# Plasmonic Twistronics:
# Discovery of Plasmonic Skyrmion Bags

Julian Schwab[1], Alexander Neuhaus[2], Pascal Dreher[2], Shai Tsesses[3], Kobi Cohen[3], Florian Mangold[1], Anant Mantha[1], Bettina Frank[1], Guy Bartal[3], Frank-J. Meyer zu Heringdorf[2], Timothy J. Davis[1,2,4], Harald Giessen[1]*

[1]4th Physics Institute, Research Center SCoPE, and Integrated Quantum Science and Technology Center, University of Stuttgart; 70569 Stuttgart, Germany.

[2]Faculty of Physics and Center for Nanointegration, Duisburg-Essen (CENIDE). University of Duisburg-Essen; 47048 Duisburg, Germany.

[3]Andrew and Erna Viterbi Department of Electrical Engineering, Technion-Israel Institute of Technology; 3200003 Haifa, Israel.

[4]School of Physics, University of Melbourne; Parkville Victoria 3010, Australia.

*Corresponding author. Email: giessen@pi4.uni-stuttgart.de

**Abstract:**

The study of van der Waals heterostructures with an interlayer twist, known as "twistronics", has been instrumental in advancing contemporary condensed matter research. Most importantly, it has underpinned the emergence of a multitude of strongly-correlated phases, many of which derive from the topology of the physical system. Here, we explore the application of the twistronics paradigm in plasmonic systems with nontrivial topology, by creating a moiré skyrmion superlattice using two superimposed plasmonic skyrmion lattices, twisted at a "magic" angle. The complex electric field distribution of the moiré skyrmion superlattice is measured using time-resolved vector microscopy, revealing that each super-cell possesses very large topological invariants and harbors a "skyrmion bag", the size of which is controllable by the twist angle and center of rotation. Our work shows how twistronics leads to a diversity of topological features in optical fields, providing a new route to locally manipulate electromagnetic field distributions, which is crucial for future structured light-matter interaction.

**Main Text:**

Moiré patterns emerge from the interference of two layers of a repeating pattern with a relative twist between them[1]. Moiré patterns were originally observed in overlays of thin fabrics, called moiré from which they derive their name, but have found diverse applications beyond textiles and artworks. In microscopy, moiré patterns enable super-resolution[2], and in photonics they provide a mechanism for creating and manipulating the spacing of optical lattices with light[3–7] and matter[8,9]. Moiré superlattices have had a profound impact in condensed matter physics, particularly in heterostructures like twisted bilayer graphene. These superlattices are responsible for such groundbreaking findings as unconventional superconductivity[10], ferroelectricity[11], and correlated insulator states[12]. The underlying mechanism for many of these exotic phases is the topology of the condensed matter system: flat moiré minibands with nontrivial topology enabling ferromagnetism and the (fractional) quantum anomalous Hall effect in Chern insulators[13,14], or charged skyrmions in bilayer graphene that facilitate superconductivity at "magic" angles[15].

Skyrmions[16] are three-dimensional topological defects on a two-dimensional plane, observed in solid materials[17–20] and in liquid crystals[21,22]. Skyrmions were recently demonstrated in optics[23–26] and particularly in plasmonic systems[27,28] through interference of surface plasmon polariton waves. However, such nontrivial topology has yet to be explored in optical moiré superlattices in general, and specifically in plasmonics.

Here, we introduce "plasmonic twistronics" that combines the tunability of moiré superlattices with the topology of skyrmion lattices to create optical fields with exotic topological features. With this method, we discover structures displaying a unique topology at "magic" twist angles that we identify as skyrmion bags[19,22,29]. These bags are multi-skyrmion textures that consist of $N$ skyrmions contained within a skyrmion boundary of opposite winding number $-1$. In addition, moiré superlattices generate super-cells with a very large topological charge, containing a skyrmion bag at their centers. We demonstrate theoretically and experimentally that the twist angle between hexagonal skyrmion lattices controls the size of skyrmion bags and their overall topological charge, showcasing the robustness of the skyrmion bags against twist angle deviations. Our experiments reveal the full vectorial electric field distributions of the plasmonic system in both time and space, which confirm that the super-cells of periodic (commensurate) moiré skyrmion lattices have characteristic higher-order topological charges. Our introduction of plasmonic twistronics provides a new paradigm for generating structured light fields with complex local topology and applications in super-resolution microscopy[30].

**Plasmonic Moiré Superlattices**

Surface plasmon polaritons (SPPs) are electromagnetic surface waves confined to the interface between metals and dielectrics. They can be excited by laser pulses incident on edges, grooves, or ridges in a metal surface to form interference patterns exhibiting different topological features in their electric field vectors[27,31] and their spin angular momentum vectors[28,32–34]. SPP waves excited from hexagonal-shaped boundaries create plasmonic skyrmion lattices, with a topology dependent upon the position of the boundaries and the resultant phase shift of the plasmons[27,35]. Plasmonic skyrmion lattices exhibit a six-fold symmetry[34], like the two-dimensional honeycomb lattices in graphene and hexagonal boron nitride (h-BN). Moiré lattices have been studied intensively in these systems, especially in twisted bilayer graphene. They are created by twisting the lower layer of the material relative to the lattice of the upper layer by the moiré twist angle $\phi$.

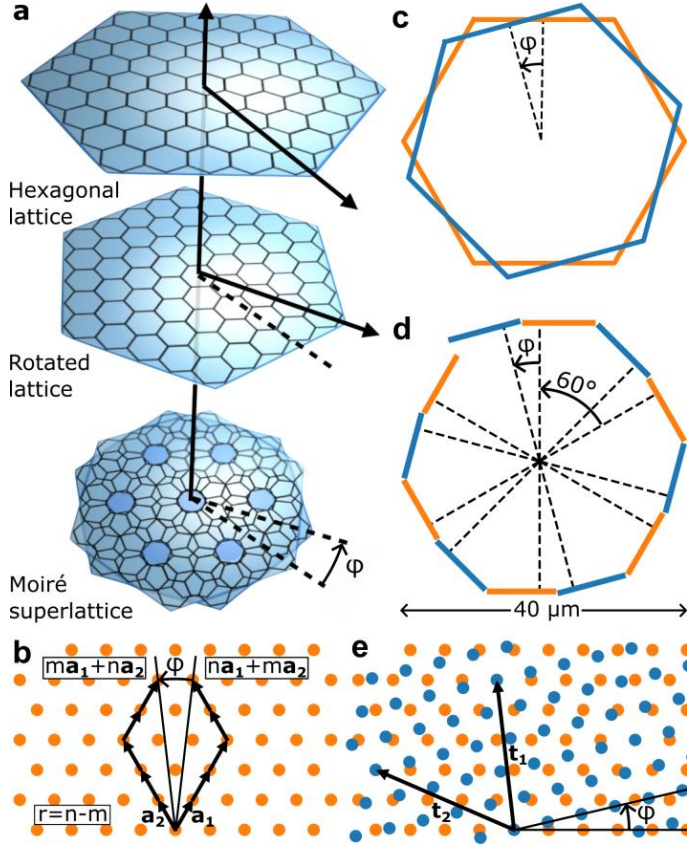

**Figure 1 | Hexagonal moiré lattices in solid state physics and plasmonics. a,** Two superimposed lattices with a relative twist create a moiré superlattice which possesses the same symmetry. **b,** Commensurate real-space lattices are created using twist angles φ that rotate an atom from position $n\boldsymbol{a_1} + m\boldsymbol{a_2}$ to $m\boldsymbol{a_1} + n\boldsymbol{a_2}$. The rotation is denoted by the integers $m$ and $r = n - m$. **c,** A skyrmion lattice is created by exciting surface plasmon polaritons along a hexagonally shaped boundary. Thus, two twisted hexagonally shaped boundaries result in a moiré skyrmion superlattice. **d,** Resulting schematic structure from which surface plasmon polaritons are emitted to create a moiré skyrmion lattice with twist angle φ. The boundaries are shifted inwards akin to an Archimedean spiral in order to counter the geometric phase arising from the circularly polarized pump pulse. **e,** Moiré superlattice and superlattice vectors $\boldsymbol{t_1}$ and $\boldsymbol{t_2}$ for the rotation φ indicated in **b**.

Superimposing the two layers results in the moiré superlattice (Fig. 1a). Moiré superlattices possess long-range aperiodic order and are hence quasicrystalline for general twist angles[36,37]. Nonetheless, for specific twist angles periodic structures can be obtained by rotating one layer over another layer, such that two lattice sites overlap. All commensurate structures in six-fold symmetry have been derived previously for graphene[1,38] and can be denoted by the integers $m$ and $n$ with the corresponding mapping

$$m\boldsymbol{a_1} + n\boldsymbol{a_2} \rightarrow n\boldsymbol{a_1} + m\boldsymbol{a_2}, \qquad m, n \in \mathbb{Z}, \tag{1}$$

as illustrated in Fig. 1b.

We create moiré skyrmion superlattices by exciting SPP waves from a boundary resembling two twisted hexagons (Fig. 1c), where each hexagon contributes a skyrmion lattice. In practice, we can combine the two hexagons to create an excitation boundary consisting of twelve segments with the interleaved and twisted hexagonal structures shown in Fig. 1d as orange and blue colors. Neighboring boundaries of the same color possess a relative angle of 60° for a hexagon whereas the relative angle between neighboring orange and blue line segments is the moiré twist angle $\phi$. Circularly polarized light is used to excite the SPPs from every boundary, resulting in spin-orbit coupling and an unwanted phase shift[35]. This phase shift is compensated by adjusting the positions (dashed lines in Fig. 1c) of the boundaries akin to an Archimedean spiral[39].

With this technique we generate a family of moiré skyrmion superlattices based on the moiré mapping equation (Eq. 1). Superlattices with small periods appear at twist angles of 21.8° ($m = 1, r = n - m = 1$), 13.2° ($m = 2, r = 1$, as in Fig. 1e), 9.4° ($m = 3, r = 1$) and 16.4° ($m = 3, r = 2$). The resulting moiré skyrmion lattices calculated for these twist angles are shown in Fig. 2 together with the corresponding superlattice vectors $\boldsymbol{t_1}$ and $\boldsymbol{t_2}$, as well as the moiré periodicity.

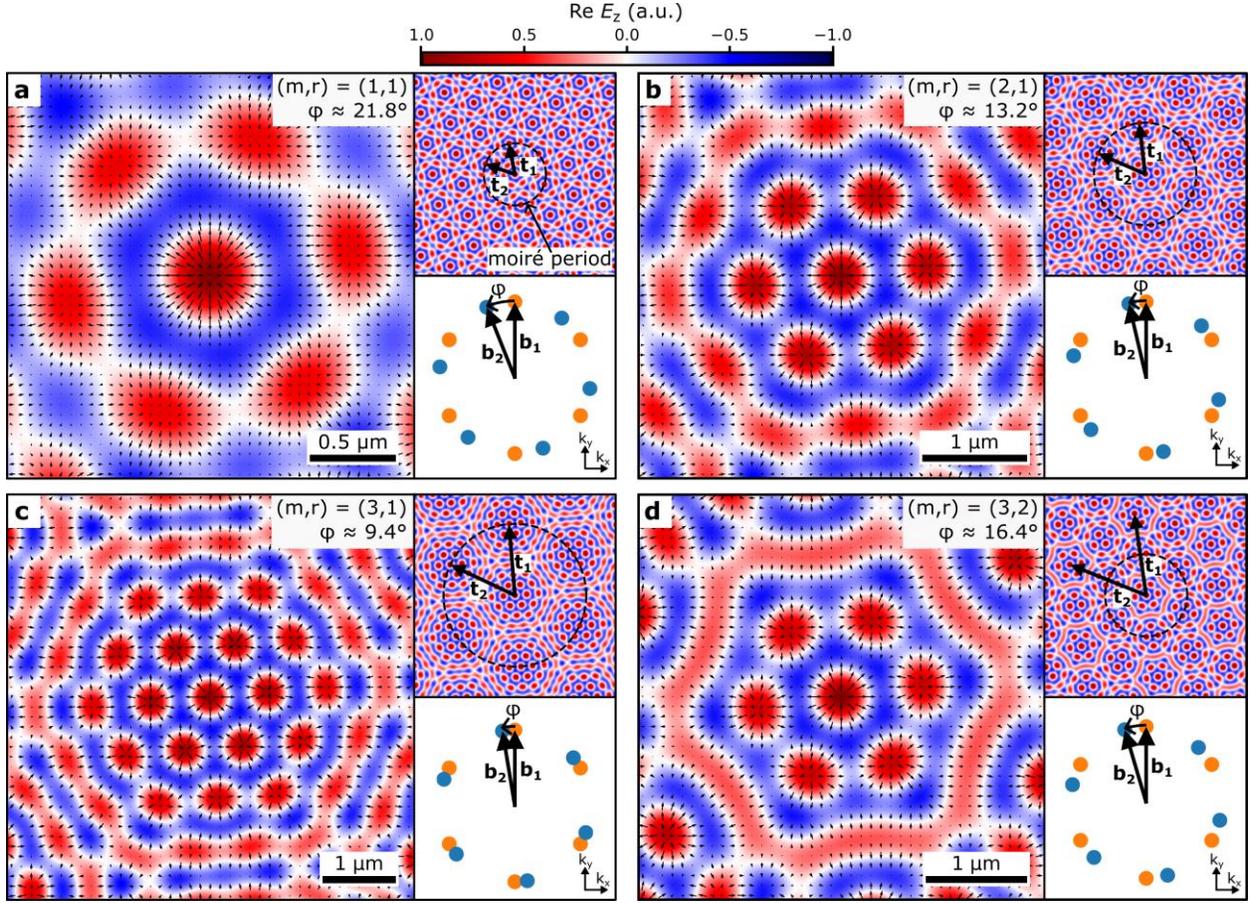

**Figure 2 | Simulated plasmonic moiré skyrmion lattices.** Simulated electric field distribution of moiré skyrmion lattices with different commensurate twist angles φ. The images on the **left** show the *z*-component of the electric field as a color plot and the in-plane components as a vector plot. The **top right** pictures display a zoomed-out view of $20 \times 20 \ \mu m^2$ together with the moiré periodicity (indicted by a circle) and the superlattice vectors $t_1$ and $t_2$. The **bottom right** images schematically represent the Fourier decomposition where the colors originate from the boundaries of Fig. 1c, d.

The electric field distributions are derived assuming infinitely long boundary lines enabling the approximation of the SPP wave as a plane wave[40]. This approximation results in an infinitely large field of view allowing the periodicity to be clearly observed (as in the top right images of Fig. 2a-d). The Fourier decomposition of the electric field (bottom right images of Fig. 2a-d) emphasizes the six-fold symmetry of the hexagonal excitation boundaries and the corresponding twist angles. Examining Fig. 2, it is clear that at the center of each super-cell exists a cluster of skyrmions surrounded by a boundary – a tell-tale sign hinting at the existence of skyrmion bags.

Although we focus in this work on moiré skyrmion superlattices formed in the SPP electric field, our findings can be readily extended to other forms of topological quasiparticles in wave systems[41,42]. In the supplement, we present numerical results of moiré skyrmion and meron superlattices[43,44] with different rotation centers, as well as moiré spin skyrmion superlattices (see Figs. S7-S13).

## Topology of Plasmonic Skyrmion Bags

The local topology of skyrmion vector fields can be characterized using the skyrmion number density

$$s = \frac{1}{4\pi} \hat{e} \cdot \left( \frac{\partial \hat{e}}{\partial x} \times \frac{\partial \hat{e}}{\partial y} \right),$$

where $\hat{e} = \boldsymbol{E}(\boldsymbol{r}, t) / ||\boldsymbol{E}(\boldsymbol{r}, t)||$ is the unit vector of the SPP electric field. The skyrmion number or topological charge is defined as $S = \int_\sigma s \, dA$, where $\sigma$ is the local region that defines the quasiparticle. The skyrmion number counts the number of times the vectors wrap around the unit sphere. It can be further divided into two additional topological numbers $S = p \cdot v$, where $p$ is the polarity, defined by the direction of the out-of-plane vector components, and $v$ is the vorticity which is determined by the direction of rotation in space of the in-plane field components[23,45].

Skyrmion bags, identified previously in liquid crystals[22] and chiral magnets[19,29], represent multi-skyrmion configurations, where $N$ skyrmions with $S = p = v = 1$ are enclosed within a larger skyrmion (Fig. 2). The surrounding skyrmion has the same vorticity but opposite polarity and therefore $S = -1$. Consequently, the total topological charge of a skyrmion bag is $S_{bag} = (N - 1)$.

Moiré lattices give rise to the formation of skyrmion bags through the superposition of two skyrmion lattices in a single plane. The twist rotates the skyrmions of one lattice around a center of rotation. When a skyrmion of the rotated lattice fills the gap between two skyrmions of the non-rotated lattice, their interference creates an elongated region with a positive $z$-component of the electric field. For suitable "magic" twist angles and centers of rotation this region forms a closed loop and hence a bag skyrmion enclosing a set of $N$ individual skyrmions (Fig. 2), thus forming a skyrmion bag with total skyrmion number $N - 1$.

When the lattice twist forms a periodic superlattice, the skyrmion bags become periodic in space and can be found in multiple super unit cells. However, the size of the skyrmion bags is always smaller than the super-cells so that skyrmion bags cannot be used to tessellate the entire 2D plane. Instead, the superlattice unit cell is rigorously defined and is associated with a well-defined topological invariant that can become very large, depending on the twist angle (supplement table S1). Its robustness is quite unique, since higher-order real-space topological charges tend to be unstable[46].

## Experimental Observation of Plasmonic Skyrmion Bags

We observe SPP skyrmion bags by measuring the spatiotemporal dynamics of the SPP electric field using time resolved two-photon photoemission (2PPE) combined with photoemission electron microscopy (PEEM)[31,47], as illustrated in Fig. 3a. In this method, a laser pulse normally incident perpendicular to the metal surface excites the conduction electrons and generates propagating SPP waves from grooves etched into a single crystalline gold flake using ion beam milling (Fig. 3b). The pump pulse has a center wavelength of 800 nm, exciting long-range surface plasmons with a wavelength of $\lambda_{SPP} = 780$ nm. The distribution of these waves is probed after a short time delay $\Delta\tau$ using a second laser pulse. The energy provided by the SPP wave and the probe pulse results in two-photon absorption and the emission of photoelectrons from the metal surface, which are imaged in an electron microscope. Since the electron yield of this process is low, many repeated measurements are performed using a femtosecond laser with 80 MHz repetition rate to obtain accurate image statistics of the electron emission. The time evolution of the SPP waves propagation and interference can be studied by repeating this process for a series of different pump-probe delays $\Delta\tau$. The resulting electron emission of the experiment is presented in Fig. 3c after application of a low-pass Fourier filter to remove a static background created by plasmo-emission. To reconstruct all vector components of the electric field, four independent pump-probe sequences are measured with different probe pulse polarizations that yield $x$ and $y$ components of the electric field. Subsequently, the out-of-plane component can be calculated using Maxwell's equations. This method allows us to reconstruct the SPP electric field components from the PEEM measurements (Fig. 3d, e) from which we can analyze the topology of the vector field.

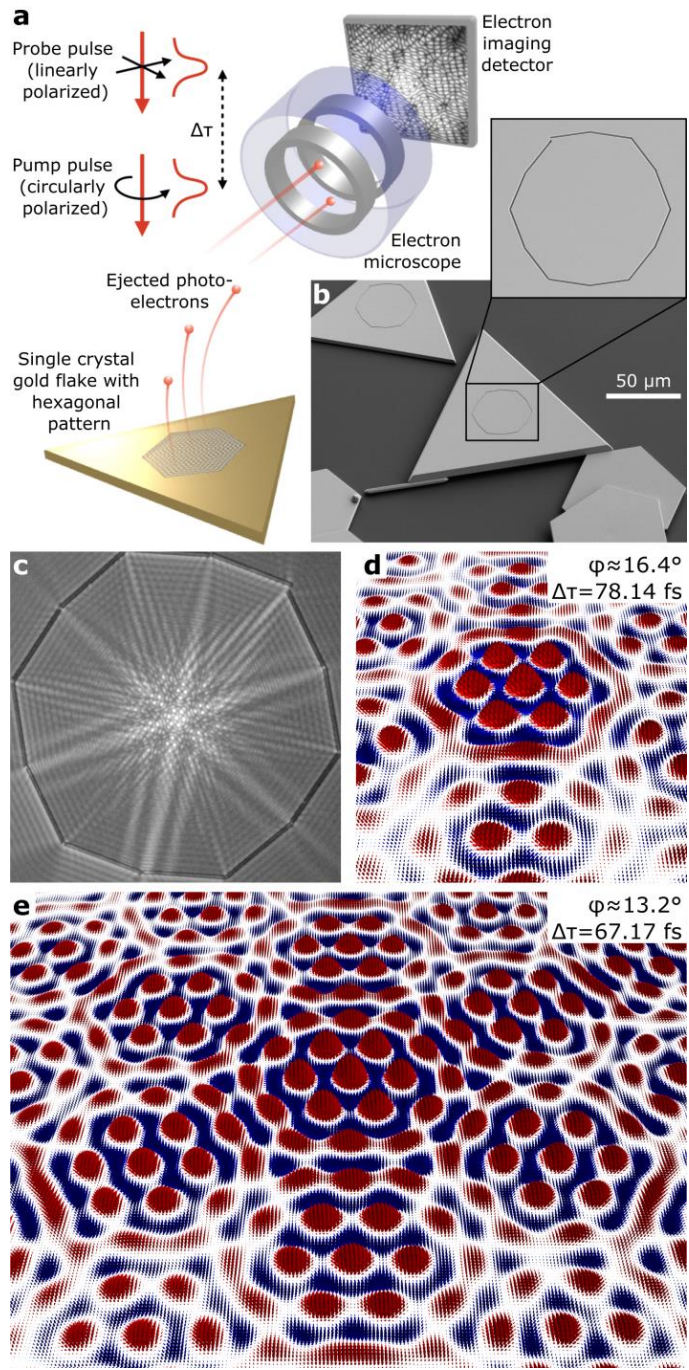

**Figure 3 | Ultrafast time-resolved vector microscopy of plasmonic skyrmion bags. a**, 2PPE-PEEM measurement method. The femtosecond laser pump-probe technique uses polarized beams combined with two-photon electron emission in an electron microscope to enable the retrieval of all vector components of the electric field of propagating surface plasmon polaritons as a function of time $\Delta\tau$. **b**, SEM images of the 16.4° structure according to Fig. 1d. The grooves are milled into single-crystalline gold flakes via ion beam lithography. **c**, Fourier filtered PEEM measurement of the plasmonic excitations. **d**, **e**, Reconstructed vector field components of surface plasmon polariton skyrmion bags at different twist angles φ. Vectors with positive (negative) $z$-component are drawn in red (blue).

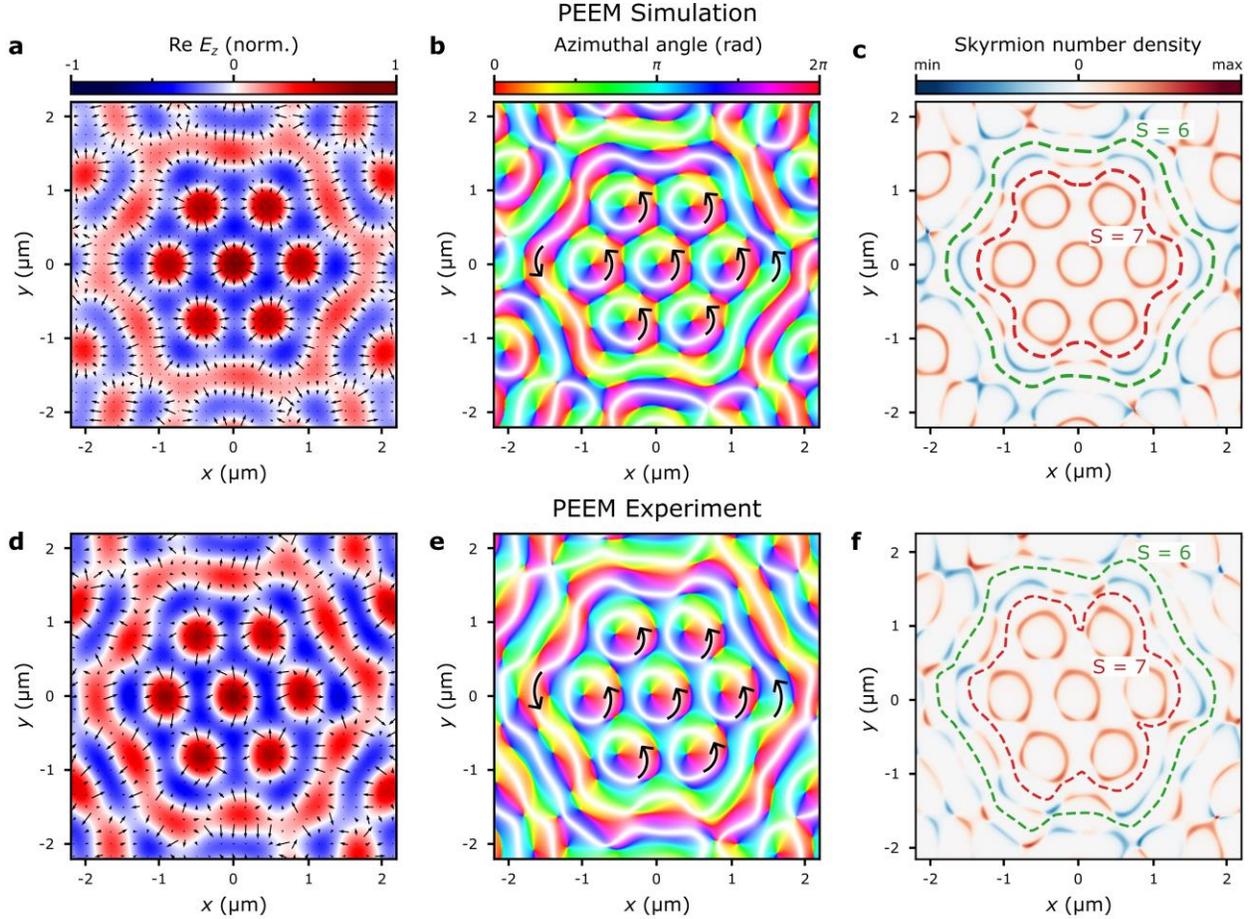

**Figure 4 | Analysis of (a-c) PEEM simulation results and (d-f) experimental results of a skyrmion bag with seven skyrmions. a, d,** Electric field distribution for the ($m = 3, r = 2$)-structure with a twist angle φ of 16.43°. The out-of-plane components is illustrated using the color plot and the in-plane component of the electric field is depicted using the vector plot. The experimental results are obtained with a delay time of $\Delta\tau = 78.14$ fs. **b, e,** Color plot of the in-plane orientation of the electric field. White areas indicate closed-loop lines, along which the electric field vectors only have an in-plane component. The in-plane orientation of the field vectors rotates by $2\pi$ along these lines of all skyrmions. Arrows indicate the direction of angle increase of the in-plane electric field direction. **c, f,** Skyrmion number density of the electric field. Integration of the skyrmion number density yields the total skyrmion number, which is calculated to $S_{bag} = 6$ for the total skyrmion bag (green dashed lines) and $S_{cluster} = 7$ for the skyrmion cluster inside the bag (red dashed lines) for the numerical and experimental results.

Our experimental method yields the temporal dynamics of the electric field vectors, revealing a periodic transition from a skyrmion bag with a topological charge of $+(N-1)$ to a skyrmion bag with a topological charge of $-(N-1)$ and back within one optical cycle. Thereby, the sign of the topological charge of both the bag skyrmion and the skyrmions inside the bag is changed analogous to the time dynamics observed in non-twisted skyrmion lattices[31]. The period of this transition is determined by the excitation pump pulse, with the length of one optical cycle being 2.67 fs for a free-space wavelength of 800 nm. A movie of the time dynamics of a skyrmion bag is provided in the supplementary movies S1-S4.

Using a structure with a twist angle of 16.4° (see Fig. 3b, c, d) a skyrmion bag is observed over multiple frames in each optical cycle over an entire measurement sequence, taken with delay times $\Delta\tau$ between 57 and 96 fs. The reconstructed field vectors are shown in Fig. 3d for a delay time of $\Delta\tau = 78.14$ fs. In Fig. 4, the same experimental results (Fig. 4d-f) are accompanied by numerical results (Fig. 4a-c), calculated using the method outlined in Ref. [40] that simulates the entire 2PPE-PEEM measurement. These data show both the out-of-plane component of the electric field vector, represented by a color plot, and the in-plane components in a vector plot. We observe remarkable agreement between experiment and theory. It is important to note that a feature can be classified as a skyrmion bag only if it satisfies the condition that the enclosed skyrmions have an individual topological charge opposite to the bag skyrmion. We examine this condition with our experimental data using the vorticity, the polarity and the local topological charges. The vorticity can be visually analyzed along closed-loop lines along which the field only has an in-plane vector component. These lines are highlighted in white (Fig. 4b, e). Since the out-of-plane fields along the boundary of the bag skyrmion are oppositely directed to those of the enclosed skyrmions (Fig. 4a, d), the polarity $p$ of the bag skyrmion is opposite to the polarity of the enclosed skyrmions. However, the azimuthal angle of the in-plane vector fields rotates in the same direction (Fig. 4b, e) so that both the bag and the enclosed skyrmions have the same vorticity $v = +1$. This leads to an opposite net skyrmion number density of the bag skyrmion (see Fig. 4c, f) compared to the enclosed skyrmions, which verifies that we indeed have created a skyrmion bag. By integrating the skyrmion number density, the topological charge for the skyrmion bag is $S_{bag} = 6$ whereas that for the skyrmions inside the bag is $S_{cluster} = 7$. As with previous approaches[28,44], the integration boundary is the path where the electric field has a maximum out-of-plane component. The numerical and experimental results fulfill the skyrmion bag requirements for a twist angle of 16.4°.

We find that the skyrmion bag appears over a range of twist angles around an optimum angle, resulting in topological robustness against twist angle rotations. The topological robustness of this structure with changes in twist angle are observed both in simulation, for a range of twist angles between 11.7 and 19.3°, and in experiment for the commensurate twist angle of 13.2° (for results see Fig. 3d and supplementary Figs. S3-S5). In general, larger twist angles result in smaller skyrmion bags that are defined over a more extended range of twist angles and therefore have increased robustness against twist angle rotations.

**More complex skyrmion bags**

Based on moiré theory, the moiré skyrmion lattice and, in turn, the skyrmion bags are limited to lattice structures with hexagonal symmetry. Thus, our method only permits skyrmion bags with specific numbers of skyrmions within the bag. For example, if the center of rotation is located at a skyrmion lattice site, as is the case in Fig. 2, only skyrmion bags with 1, 7, 19, 37, … skyrmions in the bag can be created. Larger skyrmion bags appear for lower twist angles but can be difficult to resolve in the limited field of view of the PEEM experiment. For twist angles between 21.5 and 30° a skyrmion bag with one skyrmion is formed, which is also referred to as a skyrmionium[23] (details in the supplement).

So far, only twists around the skyrmion lattice site located in the center of the structure were considered. However, the symmetry point of the twist plays an important role in the resulting topology. Skyrmion bags with different sizes are created by changing the center of rotation to other symmetry points of the lattice, as shown in the PEEM simulations in Fig. 5. This includes the set of points $P_2$ in the middle of two neighboring skyrmions at, e.g., $\boldsymbol{a_1}/2$ and the set of points $P_3$ with equal distance to three surrounding skyrmions at, e.g., $(2\boldsymbol{a_1} - \boldsymbol{a_2})/3$. With these additional rotation centers, we find other skyrmion bags of different sizes. The smallest ones are skyrmion bags with 2, 3, 4, 10, 12, and 14 skyrmions. Thus, by using the right twist angle and center of rotation it is possible to generate a tailored skyrmion bag with chosen total skyrmion number inside the moiré super cell.

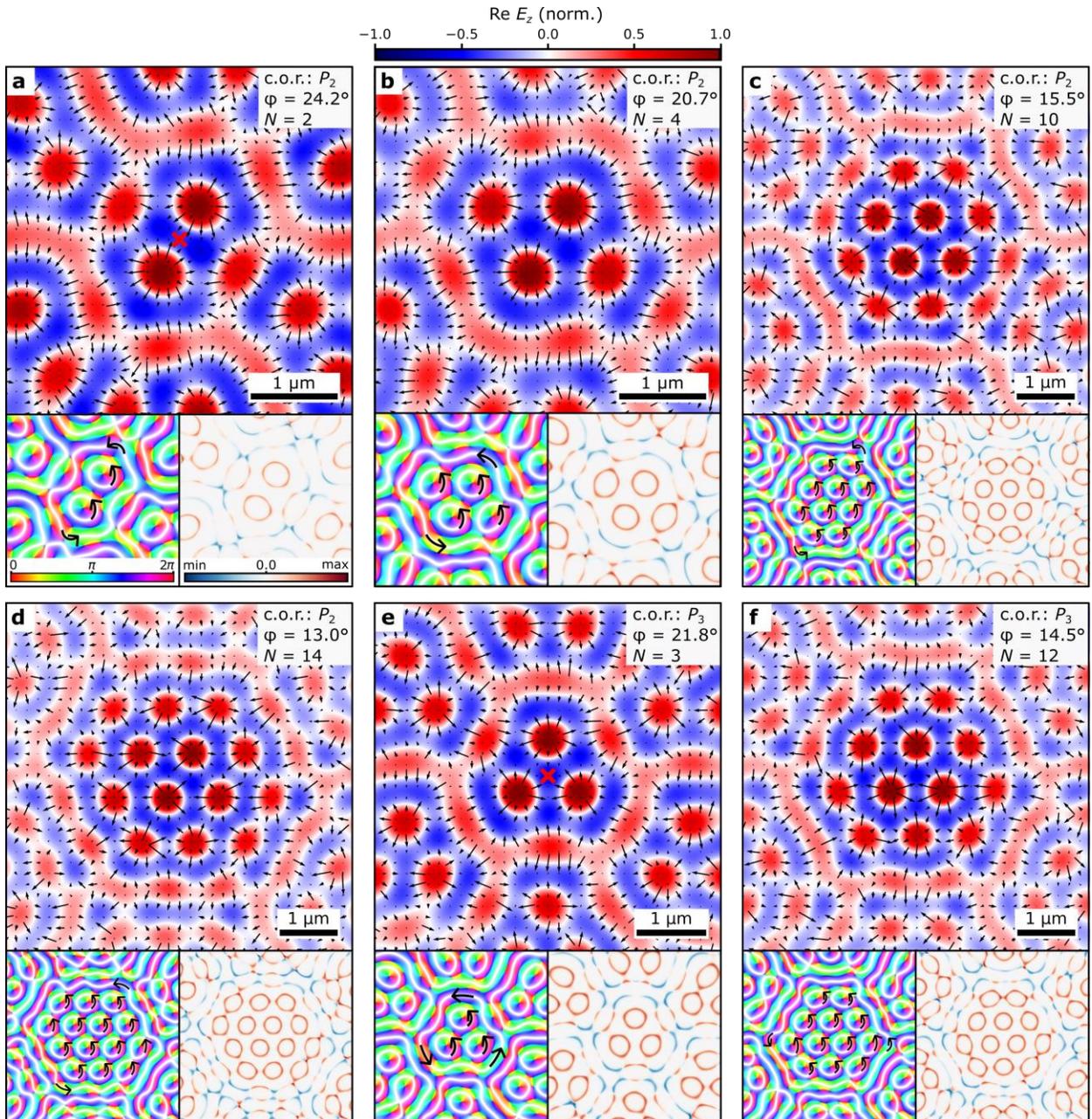

**Figure 5 | PEEM simulations of plasmonic skyrmion bags comprising (a) 2, (b) 4, (c) 10, (d) 14, (e) 3 and (f) 12 skyrmions.** The skyrmion bags depicted in a-f have $N$ skyrmions inside the bag and are obtained using the twist angle φ and the center of rotation (c.o.r.) $P_2$ at $\boldsymbol{a_1}/2$ in A-D and $P_3$ at $(2\boldsymbol{a_1} - \boldsymbol{a_2})/3$ in E, F. The rotation centers are exemplarily marked in red in a and e. The **top** image of a-f demonstrates the electric field distribution with the corresponding scale bar at the bottom right. The out-of-plane component is illustrated using the color plot and the in-plane component of the electric field is depicted using the vector plot. The **bottom left** images show a color plot of the in-plane orientation of the electric field for the same area as the image at the top. Lines, along which the electric field vectors only have an in-plane component are highlighted in white. The in-plane orientation of the field vectors rotates by $2\pi$ along these closed-loop lines of all skyrmions. Arrows indicate the direction of angle increase of the in-plane electric field direction. The **bottom right** images display the skyrmion number density of the electric field distribution in the top images.

**Outlook**

Plasmonic twistronics provides an avenue for creating electric field distributions with novel and robust topological configurations, with arbitrary large topological invariants – a longstanding goal in the budding field of topological light quasiparticles. Additional generalizations of our concept to multi-layer moiré superlattices are possible, via superimposing three or more lattices. One can also utilize other degrees of freedom, such as the local photonic spin, or lattices containing different topological quasiparticles (such as merons). In essence, our methods and results can be used to create topological states in propagating light, as well as in various other wave systems.

In general, the ability to control the topological properties of light is important for novel light-matter interaction and can be useful for applications such as spin-optics, imaging, as well as topological and quantum technologies. This is particularly important in structured light microscopy[48], for super-resolution down to the single nanometer-range[49], and structured light-matter interaction, for example non-dipolar transitions with $\Delta l \neq 1$[50] induced by light fields with non-zero local orbital angular momenta in Rydberg excitons in $Cu_2O$[51] or TMDCs[52].

# References

1. Campanera, J. M., Savini, G., Suarez-Martinez, I. & Heggie, M. I. Density functional calculations on the intricacies of Moiré patterns on graphite. *Phys Rev B Condens Matter Mater Phys* **75**, 235449 (2007).
2. Shterman, D., Gjonaj, B. & Bartal, G. Experimental Demonstration of Multi Moiré Structured Illumination Microscopy. *ACS Photonics* **5**, 1898–1902 (2018).
3. Efremidis, N. K., Sears, S., Christodoulides, D. N., Fleischer, J. W. & Segev, M. Discrete solitons in photorefractive optically induced photonic lattices. *Phys Rev E* **66**, 046602 (2002).
4. Wang, P. *et al.* Localization and delocalization of light in photonic moiré lattices. *Nature* **577**, 42–46 (2019).
5. Fu, Q. *et al.* Optical soliton formation controlled by angle twisting in photonic moiré lattices. *Nat Photonics* **14**, 663–668 (2020).
6. Wang, P. *et al.* Two-dimensional Thouless pumping of light in photonic moiré lattices. *Nat Commun* **13**, 1–8 (2022).
7. Guan, J. *et al.* Far-field coupling between moiré photonic lattices. *Nat Nanotechnol* **18**, 514–520 (2023).
8. González-Tudela, A. & Cirac, J. I. Cold atoms in twisted-bilayer optical potentials. *Phys Rev A* **100**, 053604 (2019).
9. Luo, X. W. & Zhang, C. Spin-Twisted Optical Lattices: Tunable Flat Bands and Larkin-Ovchinnikov Superfluids. *Phys Rev Lett* **126**, 103201 (2021).
10. Cao, Y. *et al.* Unconventional superconductivity in magic-angle graphene superlattices. *Nature* **556**, 43–50 (2018).
11. Zheng, Z. *et al.* Unconventional ferroelectricity in moiré heterostructures. *Nature* **588**, 71–76 (2020).
12. Cao, Y. *et al.* Correlated insulator behaviour at half-filling in magic-angle graphene superlattices. *Nature* **556**, 80–84 (2018).
13. Park, H. *et al.* Observation of fractionally quantized anomalous Hall effect. *Nature* **622**, 74–79 (2023).
14. Chen, G. *et al.* Tunable correlated Chern insulator and ferromagnetism in a moiré superlattice. *Nature* **579**, 56–61 (2020).
15. Khalaf, E., Chatterjee, S., Bultinck, N., Zaletel, M. P. & Vishwanath, A. Charged skyrmions and topological origin of superconductivity in magic-angle graphene. *Sci Adv* **7**, eabf5299 (2021).
16. Skyrme, T. H. R. A unified field theory of mesons and baryons. *Nuclear Physics* **31**, 556–569 (1962).
17. Yu, X. Z. *et al.* Real-space observation of a two-dimensional skyrmion crystal. *Nature* **465**, 901–904 (2010).
18. Mühlbauer, S. *et al.* Skyrmion lattice in a chiral magnet. *Science* **323**, 915–919 (2009).
19. Tang, J. *et al.* Magnetic skyrmion bundles and their current-driven dynamics. *Nat Nanotechnol* **16**, 1086–1091 (2021).
20. Das, S. *et al.* Observation of room-temperature polar skyrmions. *Nature* **568**, 368–372 (2019).
21. Fukuda, J. I. & Žumer, S. Quasi-two-dimensional Skyrmion lattices in a chiral nematic liquid crystal. *Nat Commun* **2**, 1–5 (2011).
22. Foster, D. *et al.* Two-dimensional skyrmion bags in liquid crystals and ferromagnets. *Nat Phys* **15**, 655–659 (2019).
23. Shen, Y. *et al.* Optical skyrmions and other topological quasiparticles of light. *Nat Photonics* **18**, 15–25 (2023).
24. Gao, S. *et al.* Paraxial skyrmionic beams. *Phys Rev A* **102**, 053513 (2020).
25. Shen, Y., Hou, Y., Papasimakis, N. & Zheludev, N. I. Supertoroidal light pulses as electromagnetic skyrmions propagating in free space. *Nat Commun* **12**, 1–9 (2021).
26. Ornelas, P., Nape, I., de Mello Koch, R. & Forbes, A. Non-local skyrmions as topologically resilient quantum entangled states of light. *Nat Photonics* **18**, 258–266 (2024).
27. Tsesses, S. *et al.* Optical skyrmion lattice in evanescent electromagnetic fields. *Science* **361**, 993–996 (2018).
28. Du, L., Yang, A., Zayats, A. V. & Yuan, X. Deep-subwavelength features of photonic skyrmions in a confined electromagnetic field with orbital angular momentum. *Nat Phys* **15**, 650–654 (2019).

29.  Yang, L. *et al.* Embedded Skyrmion Bags in Thin Films of Chiral Magnets. *Advanced Materials* 2403274 (2024).

30.  Zheludev, N. I. & Yuan, G. Optical superoscillation technologies beyond the diffraction limit. *Nature Reviews Physics* **4**, 16–32 (2021).

31.  Davis, T. J. *et al.* Ultrafast vector imaging of plasmonic skyrmion dynamics with deep subwavelength resolution. *Science* **368**, eaba6415 (2020).

32.  Spektor, G. *et al.* Revealing the subfemtosecond dynamics of orbital angular momentum in nanoplasmonic vortices. *Science* **355**, 1187–1191 (2017).

33.  Dai, Y. *et al.* Ultrafast microscopy of a twisted plasmonic spin skyrmion. *Appl Phys Rev* **9**, 11420 (2022).

34.  Lei, X. *et al.* Photonic Spin Lattices: Symmetry Constraints for Skyrmion and Meron Topologies. *Phys Rev Lett* **127**, 237403 (2021).

35.  Tsesses, S., Cohen, K., Ostrovsky, E., Gjonaj, B. & Bartal, G. Spin-Orbit Interaction of Light in Plasmonic Lattices. *Nano Lett* **19**, 4010–4016 (2019).

36.  Uri, A. *et al.* Superconductivity and strong interactions in a tunable moiré quasicrystal. *Nature* **620**, 762–767 (2023).

37.  Lifshitz, R. Quasicrystals: A matter of definition. *Found Phys* **33**, 1703–1711 (2003).

38.  Santos, J. M. B. L. dos, Peres, N. M. R. & Neto, A. H. C. Continuum model of the twisted graphene bilayer. *Phys Rev B* **86**, 155449 (2012).

39.  Gorodetski, Y., Niv, A., Kleiner, V. & Hasman, E. Observation of the spin-based plasmonic effect in nanoscale structures. *Phys Rev Lett* **101**, 043903 (2008).

40.  Davis, T. J. *et al.* Subfemtosecond and Nanometer Plasmon Dynamics with Photoelectron Microscopy: Theory and Efficient Simulations. *ACS Photonics* **4**, 2461–2469 (2017).

41.  Ge, H. *et al.* Observation of Acoustic Skyrmions. *Phys Rev Lett* **127**, 144502 (2021).

42.  Smirnova, D. A., Nori, F. & Bliokh, K. Y. Water-Wave Vortices and Skyrmions. *Phys Rev Lett* **132**, 054003 (2024).

43.  Ghosh, A. *et al.* A topological lattice of plasmonic merons. *Appl Phys Rev* **8**, 41413 (2021).

44.  Dai, Y. *et al.* Plasmonic topological quasiparticle on the nanometre and femtosecond scales. *Nature* **588**, 616–619 (2020).

45.  Göbel, B., Mertig, I. & Tretiakov, O. A. Beyond skyrmions: Review and perspectives of alternative magnetic quasiparticles. *Phys Rep* **895**, 1–28 (2021).

46.  Berry, M. V. & Dennis, M. R. Knotted and linked phase singularities in monochromatic waves. *Proceedings of the Royal Society of London. Series A: Mathematical, Physical and Engineering Sciences* **457**, 2251–2263 (2001).

47.  Meyer zu Heringdorf, F. *et al.* Spatio-temporal imaging of surface plasmon polaritons in two photon photoemission microscopy. *Proc. SPIE* **9921**, 992110–992118 (2016).

48.  Shen, Y. *et al.* Roadmap on spatiotemporal light fields. *Journal of Optics* **25**, 093001 (2023).

49.  Yang, A. *et al.* Spin-Manipulated Photonic Skyrmion-Pair for Pico-Metric Displacement Sensing. *Advanced Science* **10**, 2205249 (2023).

50.  Neubauer, A. *et al.* Spectroscopy of nanoantenna-covered $Cu_2O$: Towards enhancing quadrupole transitions in Rydberg excitons. *Phys Rev B* **106**, 165305 (2022).

51.  Kazimierczuk, T., Fröhlich, D., Scheel, S., Stolz, H. & Bayer, M. Giant Rydberg excitons in the copper oxide $Cu_2O$. *Nature* **514**, 343–347 (2014).

52.  Hill, H. M. *et al.* Observation of excitonic Rydberg states in monolayer $MoS_2$ and $WS_2$ by photoluminescence excitation spectroscopy. *Nano Lett* **15**, 2992–2997 (2015).

## Acknowledgments

The authors acknowledge support from the ERC (Complexplas, 3DPrintedoptics), DFG (SPP1391 Ultrafast Nanooptics, CRC 1242 "Non-Equilibrium Dynamics of Condensed Matter in the Time Domain" project no. 278162697-SFB 1242), BMBF (Printoptics), BW Stiftung (Spitzenforschung, Opterial), Carl-Zeiss Stiftung. T.J.D. acknowledges support from the MPI Guest Professorship Program and from the DFG (GRK2642) Photonic Quantum Engineers for a Mercator Fellowship. S.T. acknowledges support from the Adams fellowship program of the Israel Academy of Science and Humanities, the Rothschild fellowship of the Yad Hanadiv foundation, the VATAT-Quantum fellowship of the Israel Council for Higher Education, the Helen Diller Quantum Center postdoctoral fellowship, and the Viterbi fellowship of the Technion – Israel Institute of Technology.